\documentclass[conference]{IEEEtran}
\IEEEoverridecommandlockouts
\usepackage[british]{babel}
\usepackage{amsmath,epsfig}
\usepackage[utf8]{inputenc}
\usepackage[T1]{fontenc}
\usepackage{amsmath,amssymb,amsfonts}
\usepackage{graphicx}
\usepackage{textcomp}
\usepackage{pgfplots}
\usepackage{scalefnt}
\usepackage{amsmath}
\usepackage{url}

\newcommand{\diag}{\mathop{\rm diag}}

\newcommand{\nondiag}{\mathop{\rm nondiag}}
\graphicspath{{figures/}}
\def\BibTeX{{\rm B\kern-.05em{\sc i\kern-.025em b}\kern-.08em
    T\kern-.1667em\lower.7ex\hbox{E}\kern-.125emX}}
\begin{document}

\title{Joint AGC and Receiver Design for Large-Scale MU-MIMO Systems
with Coarsely Quantized Signals and C-RANs \\
\thanks{This work has been funded by CNPq, FAPERJ, CGI and CAPES.}
}
\author{\IEEEauthorblockN{T. E. B. Cunha and R. C. de Lamare}
\IEEEauthorblockA{\textit{Centre for Telecommunications Studies (CETUC)} \\
\textit{Pontifical Catholic University of Rio de Janeiro (PUC-Rio)}\\
Rio de Janeiro - 22451-900, Brazil \\
\{thiagoelias, delamare\}@cetuc.puc-rio.br} \and
\IEEEauthorblockN{T. N. Ferreira}
\IEEEauthorblockA{\textit{Engineering School} \\
\textit{Fluminense Federal University (UFF)}\\
Niterói, RJ - 24210-240, Brazil \\
tadeu\_ferreira@id.uff.br} \and \vspace{-1em}} \maketitle
\begin{abstract}
In this work, we propose the joint optimization of the automatic
gain control (AGC), which works in the remote radio heads (RHHs),
and a low-resolution aware (LRA) linear receive filter based on the
minimum mean square error (MMSE), which works on the baseband unit
(BBU) pool, for large-scale multi-user multiple-input
multiple-output (MU-MIMO) systems with coarsely quantized signals in
cloud radio access networks. We develop successive interference
cancellation receivers based on the proposed joint AGC and LRA-MMSE
(AGC-LRA-MMSE) approach. An analysis of the achievable sum rates is
also carried out. Simulations show that the proposed design has
better error rates and higher achievable rates than existing
techniques.
\end{abstract}


\section{Introduction}

In recent years, the demand for wireless communications given by the
huge number of devices connected to  networks has increased
exponentially. The expectation is that the mobile data traffic will
grow at a 47 percent compound annual growth rate (CAGR) from 2016 to
2021 \cite{cisco}. Due to the scarce spectrum, the energy
consumption, and the need for costs reduction, the development of
technologies to improve the spectral efficiency (SE) and energy
efficiency (EE) without sacrificing other communication resources
has become even more important. In this context, Cloud Radio Access
Networks (C-RANs) and large-scale MU-MIMO are key technologies to be
jointly deployed in 5G systems.

In C-RANs, the radio frequency and the processing functionalities of
the conventional base station (BS) are split into a remote radio
head (RRH) and a baseband processing  unit (BBU) pool located within
a cloud unit (CU) \cite{c_ran_zero}. Centralization allows joint
signal processing techniques that span multiple base stations,
allowing effective inter-cell interference mitigation that improves
network performance and spectral efficiency (SE)\cite{C_RAN_12}.
Large-scale MU-MIMO systems employ a large number of antennas at the
BS with the goals of enabling significant gains in both energy and
spectral efficiency \cite{ls_mimo_2}. However, the high number of
antenna elements increases considerably the hardware cost and the
energy consumption due to the presence of several analog-to-digital
(A/D) and digital-to-analog (D/A) converters
\cite{throughput_analysis}. Is well known that the energy loss in a
large-scale MU-MIMO receiver is mainly from the analog-to-digital
converter (ADC) processing unit and the digital baseband processing
unit, which are both affected by the resolution of the ADCs.
Moreover, one of the main challenges for implementing C-RANs is that
the RRHs are connected to the CU by limited-capacity links. In a
scenario where a huge amount of data is collected by large receive
antenna arrays, an often adopted approach is to compress signals
with a lower number of bits to deal with the limited-capacity
fronthaul links. Therefore, to become these technologies
commercially viable, the use of low-resolution ADCs is unavoidable
\cite{bbprec,jagcrd,1bitidd,1bitcpm}. Due to the need to minimize
the distortions arising from the quantization an AGC is used, before
the quantizers, to adjust the analog signal level to the dynamic
range of the ADC. The use of an AGC is also important in
applications where the received power varies over time, as occurs in
mobile scenarios.

In this paper, we propose an uplink framework for jointly designing
the AGCs that work in the RRHs and low-resolution aware (LRA) linear
receive filters according to the MMSE criterion, which works in the
CU of large-scale MU-MIMO with C-RANs. In order to improve the
detection performance, we propose successive interference
cancellation (SIC) receivers based on the proposed joint AGC and
LRA-MMSE (AGC-LRA-MMSE) design approach. An expression for the
achievable sum rates of the proposed AGC-LRA-MMSE with SIC receiver
is also presented. Simulations show that the BER and sum-rate
performance of AGC-LRA-MMSE with SIC using 4-6 bits outperforms
competing techniques and is close to that of a system with full
resolution.

The next section describes the uplink channel of a large-scale
MU-MIMO system with C-RANs and the model for the quantization
distortion. The expressions of the proposed AGC-LRA-MMSE receiver
design are presented in Section III. Section IV develops the sum
rate analysis. Simulations are shown and discussed in Section V.
Conclusions are given in Section VI.

\textit{Notation:} Vectors and matrices are denoted by lower and
upper case italic bold letters. The operators $(\cdot)^T$,
$(\cdot)^H$ and $\mathrm{tr}(\cdot)$ stand for transpose, Hermitian
transpose and trace of a matrix, respectively. $\mathbf{1}$ denotes
a column vector of ones and $\mathbf{I}$ denotes an identity matrix.
The operator $\mathrm{E}[\cdot]$ stands for expectation with respect
to the random variables and the operator $\odot$ corresponds to the
Hadamard product. Finally, $\diag(\mathbf{A})$ denotes a diagonal
matrix containing only the diagonal elements of $\mathbf{A}$ and
$\nondiag(\mathbf{A})=\mathbf{A}-\diag(\mathbf{A})$.

\section{System Description}

\begin{figure}[!h]
\centering
\includegraphics[scale=0.5]{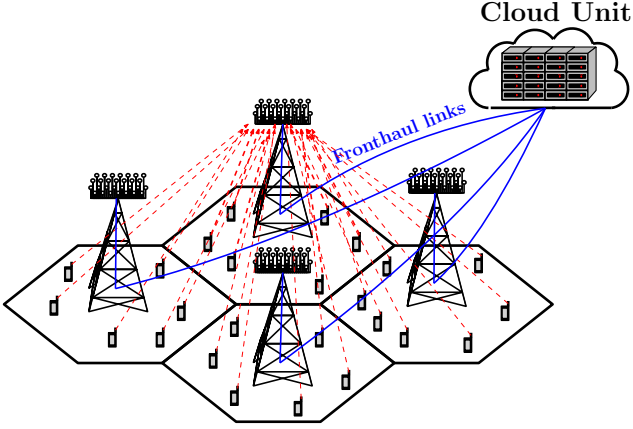}
\caption{Large-scale MU-MIMO system with CRANs.} \label{Fig:1}
\end{figure}

In this paper, we consider the uplink channel of a large-scale
MU-MIMO system with C-RANs as depicted in Fig. \ref{Fig:1}. This
system is comprised of $L$ cells where $K$ users per cell are placed
randomly, uniformly distributed over the cells. Each mobile station
(MS) and each RRH are equipped with $N_T$ transmit antennas and
$N_{R}$ receive antennas, respectively. We further assume that all
BS share the same frequency band and all users simultaneously
transmit data streams to their base stations equipped with antenna
arrays
\cite{Vantrees1,locsme,elnashar,manikas,cgbf,okspme,r19,scharf,bar-ness,pados99,
reed98,hua,goldstein,santos,qian,delamarespl07,delamaretsp,xutsa,xu&liu,
kwak,delamareccm,delamareelb,wcccm,delamarecl,delamaresp,delamaretvt,delamaretvt10,delamaretvt2011ST,
delamare_ccmmswf,jidf_echo,jidf,barc,lei09,delamare10,fa10,ccmavf,lei10,jio_ccm,
ccmavf,stap_jio,zhaocheng,zhaocheng2,arh_eusipco,arh_taes,rdrab,dcg,dce,dta_ls,
song,wljio,barc,saalt,mmimo,wence,spa,mbdf,rrmber,bfidd,did,mbthp,wlbd,baplnc,lrcc}.
Consequently, the signal transmitted by MSs in neighboring cells can
lead to intercell interference at the receivers. The
$N_{R}$-dimensional received signal vector at the RRH in the $l$-th
cell can be expressed as
\begin{eqnarray}
\mathbf{y}_{l}=\underbrace{\tilde{\mathbf{H}}_{ll}^{(k)}\mathbf{x}_l^{(k)}}_{\mathrm{\stackrel{desired}{signal}}} + \underbrace{\sum_{\stackrel{u=1}{u\neq k}}^{K}\tilde{\mathbf{H}}_{ll}^{(u)}\mathbf{x}_l^{(u)}}_{\mathrm{\stackrel{intracell}{interference}}}
+\underbrace{\sum_{\stackrel{i=1}{i\neq l}}^{L}\sum_{u=1}^{K}\tilde{\mathbf{H}}_{li}^{(u)}\mathbf{x}_i^{(u)}}_{\mathrm{\stackrel{intercell}{interference}}}\hspace{-3pt}+\hspace{-5pt}\underbrace{\mathbf{n}_{l}}_{AWGN}\hspace{-3pt},\hspace{-5pt}
\label{eq:1}
\end{eqnarray}
\noindent where $\tilde{\mathbf{H}}_{li}^{(u)} \in \mathbb{C}^{N_{R}
\times N_T}$ corresponds to the propagation matrix between the $N_T$
transmit antennas of user $u$ in the $i$th cell to the $N_{R}$
receive antennas of the RRH in the $l$th cell. Their coefficients
model independent fast and large-scale fading which are given by
$\tilde{\mathbf{H}}_{li}^{(u)}=\mathbf{H}_{li}^{(u)}\sqrt{\beta_{li}^{(u)}}.$

Each entry $h_{limn}^{(u)}$ of $\mathbf{H}_{li}^{(u)}$ represents
the fast fading coefficient from the $n$th transmit antenna of the
$u$th user in the $i$th cell to the $m$th receive antenna of the
$l$th RRH. They are assumed to be independent and identically
distributed (i.i.d) circularly symmetric complex Gaussian random
variables with zero mean and unit variance. The quantity
$\beta_{li}^{(u)}$ models both geometric attenuation and shadow
fading which are assumed to be independent, over $m$ and $n$. The
large scale coefficients can be obtained from
$\beta_{li}^{(u)}=z_{li}^{(u)}(d_{li}^{(u)}/r)^{-\gamma}$ where
$z_{li}^{(u)}$ represents the shadow fading and obeys a log-normal
distribution with standard deviation $\sigma_{\text{shadow}}$ (i.e
$10\log(z_{li}^{(u)})$ follows a Gaussian distribution with
zero-mean and standard deviation $\sigma_{\text{shadow}}$),
$d_{li}^{(u)}$ corresponds to the distance between the $u$th user in
the $i$th cell and the RRH in the $l$th cell, $r$ is the cell
radius, and $\gamma$ is the path-loss exponent \cite{large_scale_2}.
In this model, we consider the time block fading model, where the
small-scale fading channel matrix $\mathbf{H}_{li}^{(u)}$ remains
unchanged in one coherence interval and the large-scale coefficient
$\beta_{li}^{(u)}$ remains unchanged for several coherence intervals
\cite{coherence}. Let $\mathbf{x}_i^{(u)} \in \mathbb{C}^{N_T \times
1}$ as a vector of symbols transmitted by an user of the $i$th cell.
Their elements are assumed to be i.i.d circularly symmetric complex
Gaussian random variables with zero mean and unit variance.
$\mathbf{n}_l  \in \mathbb{C}^{N_{R} \times 1}$ is a vector that
denotes the additive white Gaussian noise (AWGN) at the BS in the
$l$th cell with covariance matrix
$\mathrm{E}[\mathbf{n}_l\mathbf{n}_l^H]=\sigma_n^2I_{N_{R}}$.

We can write (\ref{eq:1}) in matrix form as
$\mathbf{y}_l=\tilde{\mathbf{H}}_{l}\mathbf{x}+\mathbf{n}_l$, where
$\tilde{\mathbf{H}}_l=[\tilde{\mathbf{H}}_{l1},\tilde{\mathbf{H}}_{l2},...,\tilde{\mathbf{H}}_{lL}]$
is the $\mathbb{C}^{N_{R} \times LKN_T}$ matrix with the
coefficients of the channels between each user in the cluster and
the $N_{R}$ receive antenna of the $l$th RRH.
$\mathbf{x}=[\mathbf{x}_1^T,\mathbf{x}_2^T,...,\mathbf{x}_{L}^T]^T$
is the $\mathbb{C}^{LKN_T \times 1}$ transmit symbol vector by all
users of the cluster.

When the received signal is processed in the cloud we assume that
the BBU pool has knowledge about the received signals by all RRHs.
Therefore, the received symbol vector $\mathbf{y} \in
\mathbb{C}^{LN_{R} \times 1}$ by all RRHs of the cluster can be
written as $\mathbf{y} = \tilde{\mathbf{H}}\mathbf{x}+\mathbf{n}$,
where $\mathbf{n} \in \mathbb{C}^{LN_{R} \times 1}$ contains AWGN
samples and $\tilde{\mathbf{H}} \in \mathbb{C}^{LN_{R} \times
LKN_T}$ is the propagation matrix with the channel coefficients
between each user and each receive antennas of the cluster.

In this work, we adopt a uniform symmetric mid-riser quantizer
characterized by a set of $L=2^b$ quantization labels. The
quantization, $\mbox{Q}(y_{i,l})$, $1 \leq i \leq N_{R}$, $l \in
\{\mathcal{R},\mathcal{I}\}$, of the real or imaginary parts of the
input signal of the quantizer is a nonlinear mapping of $y_{i,l} \in
\mathcal{R}$ to a discrete set that results in additive distortion.
Its quantization rule is $r_{i,l}=\mbox{Q}(y_{i,l})=
y_{i,l}+q_{i,l}$.

The quantization factor $\rho_q^{(i,l)}$ indicates the relative
amount of quantization noise generated by a quantization process.
This factor is defined in \cite{Modified} as follows
\begin{eqnarray}
\rho_q^{(i,l)}=\frac{\mathrm{E}[q^2_{i,l}]}{r_{y_{i,l}y_{i,l}}},
\end{eqnarray}
\noindent where $r_{y_{i,l}y_{i,l}}=E[y_{i,l}^2]$ is the variance of $y_{i,l}$. This factor depends on the number of quantization bits  $b$, the quantizer type and the probability density function of $y_{i,l}$. Under the optimal design of the scalar finite resolution quantizer, the following equations hold for all $1 \leq i \leq N_{R}$, $l \in \{R,I\}$,\cite{Modified}:
    \begin{eqnarray}
    \mathrm{E}[q_{i,l}]&=&0, \\
    \mathrm{E}[r_{i,l}q_{i,l}]&=&0,\\
    \mathrm{E}[y_{i,l}q_{i,l}]&=&-\rho_q^{(i,l)}r_{y_{i,l}y_{i,l}}.
    \end{eqnarray}
Under multipath propagation conditions and for a large number of
antennas $y_{i,l}$ are approximately Gaussian distributed and  they
undergo nearly the same distortion factor $\rho_{i,l}=\rho_q$,
$\forall i$, $\forall l$. Let $q_i=q_{i,R}+jq_{i,I}$ be the complex
quantization error and under the assumption of uncorrelated real and
imaginary parts of $y_i$ we obtain $r_{q_{i}q_{i}}=\rho_q
r_{y_{i}y_{i}}$, and $r_{y_{i}q_{i}}=-\rho_q r_{y_{i}y_{i}}$. As
shown in \cite{asymptotic}, the optimal quantization step $\Delta$,
for the uniform quantizer case and for a Gaussian source, decreases
as $\sqrt{b}2^{-b}$ and $\rho_q$ is asymptotically well approximated
by $\frac{\Delta^2}{12}$.

\section{Joint AGC and LRA-MMSE Filter Design}

In this section, we present the procedure to compute the AGC
matrices and the LRA receive filter. We initialize the algorithm by
computing the modified MMSE receiver $\mathbf{W}_l$, that takes into
account quantization, for each of the cluster cells. In order to
obtain the optimal AGC coefficients, the derivative of the cost
function that takes into account both the presence of the AGC and
the estimated signal is computed in an alternating fashion
\cite{joint}. Then, to detect the desired symbols in the cloud, an
AGC-LRA-MMSE receive filter with SIC (AGC-LRA-MMSE-SIC), which
considers the quantization effects and the presence of the AGCs, is
derived.

\subsection{Low Resolution Aware (LRA-MMSE) Receive Filter}

The received signal after the quantizer is expressed, with the
\textit{Bussgang} decomposition \cite{bussgang}, as a linear model
$\mathbf{r}_l=\mathbf{y}_l+\mathbf{q}_l$. Then, the linear receive
filter $\mathbf{W}_l  \in \mathbb{C}^{KN_T \times N_{R}}$ that
minimizes the MSE
\begin{eqnarray}
\varepsilon =E[||\mathbf{x}_l-\hat{\mathbf{x}}_l||^2_2]=E[||\mathbf{x}_l-\mathbf{W}_l\mathbf{r}_l||^2_2],
\end{eqnarray}
\noindent is given by
\begin{eqnarray}
\mathbf{W}_l=\mathbf{R}_{x_lr_l}\mathbf{R}_{r_lr_l}^{-1}.
\label{eq:wiener}
\end{eqnarray}

\noindent where the cross-correlation matrix $\mathbf{R}_{x_lr_l} \in \mathbb{C}^{KN_T \times N_{R}}$, and the autocorrelation matrix $\mathbf{R}_{r_lr_l} \in \mathbb{C}^{N_{R} \times N_{R}}$ are expressed as
\begin{eqnarray}
\mathbf{R}_{x_lr_l}&=&\mathbf{R}_{x_ly_l}+\mathbf{R}_{x_lq_l},
\label{eq:Rxr} \\
\mathbf{R}_{r_lr_l}&=&\mathbf{R}_{y_ly_l}+\mathbf{R}_{y_lq_l}+\mathbf{R}_{y_lq_l}^H+\mathbf{R}_{q_lq_l}.
\label{eq:Rrr}
\end{eqnarray}

We can get the autocorrelation matrix $\mathbf{R}_{y_ly_l} \in
\mathbb{C}^{N_{R} \times N_{R}}$ and the cross-correlation matrix
$\mathbf{R}_{x_ly_l} \in \mathbb{C}^{KN_T \times N_{R}}$ directly
from the MU-MIMO model as
\begin{align}
\mathbf{R}_{y_ly_l}&=\tilde{\mathbf{H}}_l\mathbf{R}_{xx}\tilde{\mathbf{H}}_l^H+\mathbf{R}_{n_ln_l}, \\
\mathbf{R}_{x_ly_l}&=\mathbf{R}_{x_lx_l}\tilde{\mathbf{H}}_{ll}^H.
\end{align}
To compute equations (\ref{eq:Rrr}) and (\ref{eq:Rxr}) we need to
obtain the covariance matrices $\mathbf{R}_{x_lq_l}$,
$\mathbf{R}_{y_lq_l}$ and $\mathbf{R}_{q_lq_l}$ as a function of the
channel parameters and the distortion factor $\rho_q$. The procedure
to obtain these matrices was developed in \cite{Modified} and we
employ the expressions:
\begin{align}
\mathbf{R}_{x_lq_l} &= -\rho_q\mathbf{R}_{x_ly_l}, \label{eq:Rxq} \\
\mathbf{R}_{y_lq_l} &\approx -\rho_q\mathbf{R}_{y_ly_l}, \label{eq:Ryq}\\
\mathbf{R}_{q_lq_l} &\approx \rho_q (\mathbf{R}_{y_ly_l}-(1-\rho_q)\nondiag(\mathbf{R}_{y_ly_l})).  \label{eq:Rqq}
\end{align}
Substituting (\ref{eq:Rxq}) in (\ref{eq:Rxr}) and (\ref{eq:Ryq}) and
(\ref{eq:Rqq}) in  (\ref{eq:Rrr}), we get
\begin{eqnarray}
\mathbf{R}_{x_lr_l}&=&(1-\rho_q)\mathbf{R}_{x_ly_l}, \label{Rxr}\\
\mathbf{R}_{r_lr_l}&=&(1-\rho_q)(\mathbf{R}_{y_ly_l}-\rho_q\nondiag(\mathbf{R}_{y_ly_l})).\label{Rrr}
\end{eqnarray}
Substituting (\ref{Rxr}) and (\ref{Rrr}) in (\ref{eq:wiener}) we
obtain the expression of the LRA-MMSE receive filter for a coarsely
quantized MU-MIMO system for the $l$th cell:
    \begin{eqnarray}
    \mathbf{W}_l \approx \mathbf{R}_{x_ly_l}(\mathbf{R}_{y_ly_l}-\rho_q\nondiag(\mathbf{R}_{y_ly_l}))^{-1}.
    \end{eqnarray}

\subsection{AGC Design}
To obtain the optimum AGC coefficients we compute the derivative of
the MSE cost function with respect to the matrix $\mathbf{G}_l$
keeping $\mathbf{W}_l$ fixed. Therefore, we will have a
initialization using the LRA-MMSE linear filter $\mathbf{W}_l$
previously computed. After that we will obtain $\mathbf{G}_l$.
Consider $\mathbf{G}_l  \in \mathbb{R}^{N_{R} \times N_{R}}$ as a
diagonal matrix and $\mathbf{g}_l \in \mathbb{R}^{N_{R} \times 1}$ a
vector with the diagonal coefficients of $\mathbf{G}_l$. We can
write $\mathbf{G}_l= \diag(\mathbf{g}_l)$. The MSE cost function can
be written as
\begin{eqnarray}
\varepsilon =E[||\mathbf{x}_l-\mathbf{W}_l(\alpha\diag(\mathbf{g}_l)\mathbf{y}_l+\mathbf{q}_l)||^2_2],
\end{eqnarray}
    \noindent where $\alpha$ corresponds to the clipping factor of the AGC. This factor is a commonly used rule to adjust the amplitude of the received signal in order to minimize the overload distortion. To obtain the optimum $\mathbf{G}_l$ matrix we compute the derivative of the MSE cost function with respect to $\diag(\mathbf{g}_l)$, equate the derivatives to zero, and then solve for $\mathbf{g}_l$. After this, we get
     \begin{align}
     \mathbf{g}_l&=[(\mathbf{W}_l^T\mathbf{W}_l^*) \odot \mathbf{R}_{y_ly_l}+(\mathbf{W}_l^H\mathbf{W}_l) \odot \mathbf{R}_{y_ly_l}^T]^{-1} \nonumber \\
     &\cdot\frac{2}{\alpha}(Re([(\mathbf{R}_{x_ly_l}^T\odot\mathbf{W}_l^H)\mathbf{1}])
     - Re([(\mathbf{W}_l^T\odot(\mathbf{R}_{y_lq_l}\mathbf{W}_l^H))\mathbf{1}])),
     \nonumber
     \end{align}
     \noindent where $\mathbf{1}$ is a $N_{R} \times 1$ vector of ones. The optimum AGC matrix can be written as $\mathbf{G}_l=\diag(\mathbf{g}_l)$. Based on the received signal power, the clipping factor $\alpha$ can be obtained from
     \begin{eqnarray}
     \alpha=\beta \cdot\sqrt{\frac{\mathrm{tr}(\mathbf{R}_{y_ly_l}+\mathbf{R}_{y_lq_l}+\mathbf{R}_{y_lq_l}^H+\mathbf{R}_{q_lq_l})}{N_{R}}},
     \end{eqnarray}
\noindent where $\beta$ is a calibration factor. To ensure an
optimized performance, the value of $\beta$ was set to
$\frac{\sqrt{b}}{2}$ in simulations, which corresponds to the
modulus of the last quantizer label.

Before quantizing, the received signals are treated by independent
AGCs. After that, the quantized signals are sent to the CU and, at
the same time, the respective AGC coefficients are also sent. In the
BBU pool an $\mathbb{R}^{LN_{R} \times LN_{R}}$ AGC matrix that
contains all AGC coefficients of the cluster can be organized as
$\mathbf{G}=\diag([\mathbf{g}_1^T,...,\mathbf{g}_L^T]^T)$.

\subsection{AGC-LRA-MMSE-SIC receiver in the CU}

SIC detectors can outperform linear detectors and achieve the
sum-capacity in the uplink of MU-MIMO systems. At each stage, one
data stream is decoded and its contribution is removed from the
received signal. To minimize error propagation, data streams are
ranked based on reliability measures such as log-likelihood ratios
or channel powers \cite{sic_1,mbdf}. At the $a$-th stage the
received signal vector, $\mathbf{y}^{(a)} \in \mathbb{C}^{LN_{R}
\times 1}$ in the CU, is given by
    $\mathbf{y}^{(a)}=\mathbf{y}^{(1)}$, if a=1, and
    $\mathbf{y}^{(1)}-\displaystyle \sum_{j=1}^{a-1}\tilde{\mathbf{h}}^{\mathbf{\Phi}(j)}\hat{x}^{\mathbf{\Phi}(j)}$, if $2\leqslant a \leqslant LKN_T$, where $\hat{x}^{\mathbf{\Phi}(j)}$ is the symbol estimated at the $j$-th stage prior to the $a$-th stage and $\tilde{\mathbf{h}}^{\mathbf{\Phi}(j)} \in \mathbb{C}^{LN_{R} \times 1}$ is the $\mathbf{\Phi}(j)$-th column of  $\tilde{\mathbf{H}}$. In this notation, $\mathbf{\Phi}$ corresponds to the ranking vector, whose entries indicates what symbol is detected at each stage. After detection, the corresponding column $\tilde{\mathbf{h}}^{\mathbf{\Phi}(a)}$ from the channel matrix $\tilde{\mathbf{H}}^{(a)} \in \mathbb{C}^{LN_{R} \times (LKN_T-a+1)}$ is nullified and a new LRA-MMSE receive filter is computed for the next stage. The quantized received signal vector $\mathbf{r}^{(a)} \in \mathbb{C}^{LN_{R} \times 1}$, in the $a$-th stage, is given by
\begin{eqnarray}
\mathbf{r}^{(a)} = Q(\mathbf{G}\mathbf{y}^{(a)}) =  \mathbf{G}(\tilde{\mathbf{H}}^{(a)}\mathbf{x}^{(a)}+\mathbf{n})+\mathbf{q}^{(a)}.
\end{eqnarray}
 To compute the LRA-MMSE linear receive filter we use the \textit{Wiener-Hopf} equations $\mathbf{W}_{LRA}^{(a)}=\mathbf{R}_{xr}^{(a)}(\mathbf{R}_{rr}^{(a)})^{-1}$, where the the cross-correlation matrix $\mathbf{R}_{xr}^{(a)} \in \mathbb{C}^{(LKN_T-a+1) \times LN_{R}}$ and autocorrelation matrix $\mathbf{R}_{rr}^{(a)} \in \mathbb{C}^{LN_{R} \times LN_{R}}$ are given by
\begin{eqnarray}
\mathbf{R}_{xr}^{(a)}&=&\mathbf{R}_{xy}^{(a)}\mathbf{G}+\mathbf{R}_{xq}^{(a)}, \label{eq:Rxr_wiener2} \\
\mathbf{R}_{rr}^{(a)}&=&\mathbf{G}\mathbf{R}_{yy}^{(a)}\mathbf{G}+\mathbf{G}\mathbf{R}_{yq}^{(a)}+(\mathbf{R}_{yq}^{(a)})^H\mathbf{G}+\mathbf{R}_{qq}^{(a)}.\label{eq:Rrr_wiener2}
\end{eqnarray}

\section{Sum Rate Analysis}

The uplink sum rate of the AGC-LRA-MMSE-SIC receiver in a system
with $LKN_T$ interfering layers is equal to the sum of the
achi\-e\-va\-ble rate of the $a$-th stream after the AGC-LRA-MMSE
receiver, and the achievable rate of the reduced size $(LKN_T-a)
\times LN_{R}$ MIMO system after removal of the $a$-th stream, given
by
\begin{align}
\mathcal{R}_{sum}=\sum_{a=1}^{MKN_T} E\left[\log_2\left(1+\frac{\Upsilon^{\mathbf{\Phi}(a)}}{\Gamma^{\mathbf{\Phi}(a)}}\right)\right],
\label{eq:sum_sic}
\end{align}
\noindent where $\Upsilon^{\mathbf{\Phi}(a)}$ is the desired signal power and $\Gamma^{\mathbf{\Phi}(a)}$ is the interference plus noise  power. The expectation is taken over the channel coefficients.
    In the $a$-th stage, the estimated symbol is given by
\vspace{-8pt}
\begin{align}
\hat{x}_l^{\mathbf{\Phi}(a)}=&\mathbf{w}_{LRA,l}^{\mathbf{\Phi}(a)}\mathbf{G}\tilde{\mathbf{h}}_{l}^{\mathbf{\Phi}(a)}x_l^{\mathbf{\Phi}(a)}+\sum_{\stackrel{u=1}{u\neq \mathbf{\Phi}(a)}}^{KN_T}\mathbf{w}_{LRA,l}^{\mathbf{\Phi}(a)}\mathbf{G}\tilde{\mathbf{h}}_{l}^{(u)}x_l^{(u)} \nonumber \\
&+\sum_{\stackrel{j=1}{j\neq l}}^{L}\sum_{u=1}^{KN_T}\mathbf{w}_{LRA,l}^{\mathbf{\Phi}(a)}\mathbf{G}\tilde{\mathbf{h}}_{j}^{(u)}x_j^{(u)}+\mathbf{w}_{LRA,l}^{\mathbf{\Phi}(a)}\mathbf{G}\mathbf{n} \nonumber \\
&+\mathbf{w}_{LRA,l}^{\mathbf{\Phi}(a)}\mathbf{q}, \hspace{15pt} \{x_l^{\mathbf{\Phi}(a)},x_l^{(u)},x_j^{(u)}\} \not\subset \mathbf{\Omega},
\end{align}
\noindent where $\mathbf{\Omega}$ is a set of symbols estimated at
prior stages. The coefficients of the receive filter
$\mathbf{w}_{LRA,l}^{\mathbf{\Phi}(a)}$ are obtained from the
$\mathbf{\Phi}(a)$-th row of the filter matrix
$\mathbf{W}_{LRA}^{(a)}$. Given a channel realization
$\tilde{\mathbf{H}}$, the desired signal power is computed by

\begin{eqnarray}
\Upsilon^{\mathbf{\Phi}(a)}=\sigma_x^2(\mathbf{w}_{LRA,l}^{\mathbf{\Phi}(a)}\mathbf{G}\tilde{\mathbf{h}}_{l}^{\mathbf{\Phi}(a)})(\mathbf{w}_{LRA,l}^{\mathbf{\Phi}(a)}\mathbf{G}\tilde{\mathbf{h}}_{l}^{\mathbf{\Phi}(a)})^H,
\label{eq:Upsilon_sic}
\end{eqnarray}

    \noindent where $\tilde{\mathbf{h}}^{\mathbf{\Phi}(a)}_l$ is the $\mathbf{\Phi}(a)$-th column of  $\tilde{\mathbf{H}}^{(a)}$. Then, $\tilde{\mathbf{h}}_{l}^{\mathbf{\Phi}(a)}$ becomes null and the interference plus noise power is
    \vspace{-8pt}
\begin{align}
\Gamma^{\mathbf{\Phi}(a)}=&\sigma_x^2(\mathbf{w}_{LRA,l}^{\mathbf{\Phi}(a)}\mathbf{G}\tilde{\mathbf{H}}^{(a)})(\mathbf{w}_{LRA,l}^{\mathbf{\Phi}(a)}\mathbf{G}\tilde{\mathbf{H}}^{(a)})^H  \nonumber \\
&-\rho_q \sigma_x^2[(\mathbf{w}_{LRA,l}^{\mathbf{\Phi}(a)}\tilde{\mathbf{H}}_{l}^{(a)})(\mathbf{w}_{LRA,l}^{\mathbf{\Phi}(a)}\mathbf{G}\tilde{\mathbf{H}}_{l}^{(a)})^H \nonumber \\
&+(\mathbf{w}_{LRA,l}^{\mathbf{\Phi}(a)}\mathbf{G}\tilde{\mathbf{H}}_{l}^{(a)})(\mathbf{w}_{LRA,l}^{\mathbf{\Phi}(a)}\tilde{\mathbf{H}}_{l}^{(a)})^H \nonumber \\
&+\sum_{\stackrel{j=1}{j\neq l}}^{L}(\mathbf{w}_{LRA,l}^{\mathbf{\Phi}(a)}\tilde{\mathbf{H}}_{j}^{(a)})(\mathbf{w}_{LRA,l}^{\mathbf{\Phi}(a)}\mathbf{G}\tilde{\mathbf{H}}_{j}^{(a)})^H  \nonumber \\
&+\sum_{\stackrel{j=1}{j\neq l}}^{L}(\mathbf{w}_{LRA,l}^{\mathbf{\Phi}(a)}\mathbf{G}\tilde{\mathbf{H}}_{j}^{(a)})(\mathbf{w}_{LRA,l}^{\mathbf{\Phi}(a)}\tilde{\mathbf{H}}_{j}^{(a)})^H]  \nonumber \\
&+\sigma_n^2(\mathbf{w}_{LRA,l}^{\mathbf{\Phi}(a)}\mathbf{G})(\mathbf{w}_{LRA,l}^{\mathbf{\Phi}(a)}\mathbf{G})^H\nonumber \\
&-\rho_q\sigma_n^2[(\mathbf{w}_{LRA,l}^{\mathbf{\Phi}(a)})(\mathbf{w}_{LRA,l}^{\mathbf{\Phi}(a)}\mathbf{G})^H\nonumber \\
&+(\mathbf{w}_{LRA,l}^{\mathbf{\Phi}(a)}\mathbf{G})(\mathbf{w}_{LRA,l}^{\mathbf{\Phi}(a)})^H]\nonumber \\
&+\mathbf{w}_{LRA}^{\mathbf{\Phi}(a)}\rho_q(\mathbf{R}_{yy}-(1-\rho_q) \nondiag(\mathbf{R}_{yy}))\mathbf{w}_{LRA}^{\mathbf{\Phi}(a)H},
\label{eq:gamma_sic}
\end{align}

\noindent where $\tilde{\mathbf{H}}_{l}^{(a)}$ is the channel matrix
between the users in the $l$-th cell and all receive antennas.
Substituting (\ref{eq:Upsilon_sic}) and (\ref{eq:gamma_sic}) in
(\ref{eq:sum_sic}) we get the achievable sum rate
$\mathcal{R}_{sum}$ of the system.

\section{Results}

In this section, we evaluate the performance of the proposed
AGC-LRA-MMSE-SIC design. We consider uplink C-RANs composed of 4
RRHs connected to the CU. The RRHs share the same frequency band.
Each cell contains in the covered area one centralized RRH equipped
with 64 receive antennas and a total of $K=8$ users equipped with
$N_T=2$ transmit antennas each. The channel model in the simulations
includes fast fading, geometric attenuation, and log-normal shadow
fading. The small-scale fading is modeled by a Rayleigh channel
whose coefficients are i.i.d complex Gaussian random variables with
zero-mean and unit variance. The large-scale fading coefficients are
obtained by
$\beta_{li}^{(u)}=z_{li}^{(u)}(d_{li}^{(u)}/r)^{-\gamma}$, where the
path-loss exponent is $\gamma=3.7$, and the shadow-fading standard
deviation is $\sigma_{shadow}=8.0$ dB. We consider a cell radius of
$r_c=1000$ meters and the users are randomly distributed in a
covered area between a cell-hole radius of $r_h=200$ meters and the
cell edge. For each channel realization, the users transmit data
packets with 100 symbols using QPSK modulation. SIC receivers employ
channel norm-based ordering. The simulated bit error ratio (BER) is
averaged over the transmission of 100 packets, by each transmit
antenna of each user. In each RRH the received signals are treated
by independent AGCs and then quantized by $b$-bit resolution uniform
quantizers before signal transmission to the BBU.

\begin{figure}[h]
\centering
\includegraphics[scale=0.5]{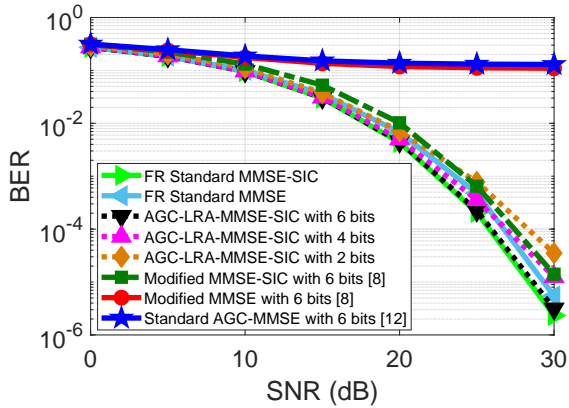}
\vspace{-1pt} \caption{Uncoded BER performance comparison with QPSK
modulation considering perfect CSIR.} \label{fig:ber_QPSK}
\end{figure}

In Fig. \ref{fig:ber_QPSK} we investigate the BER performance  gain
achieved by the proposed AGC-LRA-MMSE receiver design with SIC
detection scheme (AGC-LRA-MMSE-SIC) when users transmit QPSK
symbols. To investigate this we consider the Modified MMSE receiver
presented in \cite{Modified} and the standard AGC from \cite{b2}
with the standard MMSE receiver. For a fair comparison we also
employ \cite{Modified} with the SIC detection scheme. Through this
result is possible to see that in a system whose signals are
quantized with $6$ bits, the proposed AGC-LRA-MMSE-SIC approach
achieves a very close performance to the performance achieved by the
Full-Resolution (FR) standard MMSE-SIC receiver in a system with
signals quantized with 16 bits. Moreover, the proposed
AGC-LRA-MMSE-SIC detection scheme has a significantly better
performance than existing techniques.

\begin{figure}[h]
\centering
\includegraphics[scale=0.5]{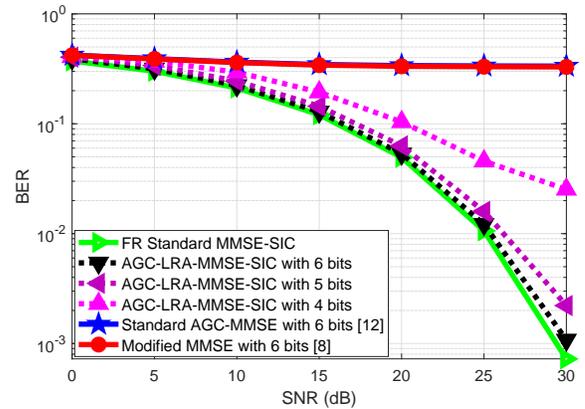}
\vspace{-1pt} \caption{Uncoded BER performance comparison with
16-QAM modulation.} \label{fig:ber_16QAM}
\end{figure}

Increasing the modulation order we evaluate in Fig.
\ref{fig:ber_16QAM} the BER performance achieved by the proposed
scheme in a scenario whose users transmits 16-QAM modulation
symbols. A comparison between Fig. \ref{fig:ber_16QAM} and Fig.
\ref{fig:ber_QPSK} shows a significant performance loss due to the
higher modulation order. In Fig. \ref{fig:ber_16QAM} the BER
performance achieved by the AGC-LRA-MMSE-SIC scheme when signals are
quantized with $5$ or $6$ is close to the FR Standard MMSE-SIC
receiver. We can also see that the proposed scheme achieves a better
performance than existing techniques. This result indicates that the
proposed AGC-LRA-MMSE-SIC scheme can improve the BER performance
even when users transmit symbols of a higher modulation order.

\begin{figure}[h]
\centering
\includegraphics[scale=0.5]{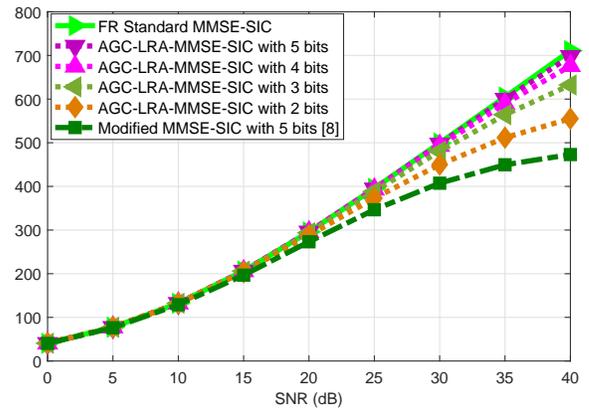}
\vspace{-1pt} \caption{Achievable sum rates of the AGC-LRA-MMSE-SIC
scheme.} \label{fig:sum_rate_sic}
\end{figure}

Fig. \ref{fig:sum_rate_sic} compares the achievable sum rates by the
proposed AGC-LRA-MMSE-SIC receiver and the sum rates achieved by the
FR Standard MMSE-SIC receiver. We can see that, the proposed
receiver achieves a sum rate similar to the FR Standard MMSE-SIC
receiver, even in a system whose signals are quantized with 5 bits.

The improved performance achieved by the proposed AGC-LRA-MMSE-SIC
design is due the optimizations of the AGC that adjusts the analog
signal level to the dynamic range of the ADC, the receiver that
takes into account the quantization distortion and the additional
gain achieved by SIC detection.

\section{Conclusions}

We have proposed the joint design of the AGC and LRA-MMSE receive
filters for coarsely quantized large-scale MU-MIMO systems in
C-RANs. The proposed AGC-LRA-MMSE-SIC receiver design outperforms
competing low-resolution receiver designs and achieves a performance
very close to that achieved by the FR Standard MMSE-SIC receiver in
terms of BER and sum rates. 

\end{document}